# The case for space environmentalism

*A.Lawrence, M.L.Rawls, M.Jah, A.Boley, F. Di Vruno, S.Garrington, M.Kramer, S.Lawler, J.Lowenthal, J.McDowell, M.McCaughrean.*

## Abstract

The shell bound by the Karman line at a height of ~80–100km above the Earth's surface, and Geosynchronous Orbit, at ~36,000km, is defined as the orbital space surrounding the Earth. It is within this region, and especially in Low Earth Orbit (LEO), where environmental issues are becoming urgent because of the rapid growth of the anthropogenic space object population, including satellite 'mega-constellations'. In this Perspective, we summarise the case that the orbital space around the Earth should be considered an additional ecosystem, and so subject to the same care and concerns and the same broad regulations as, for example, the oceans and the atmosphere. We rely on the orbital space environment by looking through it as well as by working within it. Hence, we should consider damage to professional astronomy, public stargazing and the cultural importance of the sky, as well as the sustainability of commercial, civic and military activity in space. Damage to the orbital space environment has problematic features in common with other types of environmental issue. First, the observed and predicted damage is incremental and complex, with many contributors. Second, whether or not space is formally and legally seen as a global commons, the growing commercial exploitation of what may appear a 'free' resource is in fact externalising the true costs.

## Main

This article has its origin in an Amicus Brief [1] submitted to the US Court of Appeal in August 2021, in support of an appeal made by several organisations against a specific order made by the US Federal Communications Commission (FCC). That order granted license amendments for SpaceX Starlink satellites. As we write, the appeal process is still underway, but all submissions to the Court have been made, so it is appropriate to make the material public. Since constructing the Amicus Brief, similar very general environmental arguments have been made in an article by L. Miraux [2].

# Orbital space and its regions

Most anthropogenic space activity is between altitudes of 100km to 36,000km. For the purposes of this article, we refer to this as "near Earth orbital space", traditionally classified in three broad regions.

**Low Earth Orbit** (LEO) is generally understood to be at altitudes of 100 to 2,000km, with many **anthropogenic space objects** (ASOs) at around 500km. LEO has traditionally been dominated by scientific, earth observation and military missions, with some communications systems. The orbital period at these altitudes is around 90-120 minutes, and so, seen from Earth, any one satellite moves across the entire sky in a few minutes. Orbits in this region decay due to atmospheric drag, but the timescale varies significantly from a few months at the lowest altitudes to hundreds of years above ~1,200km.

**Medium Earth Orbit** (MEO) is at altitudes around 20,000 km, within a broad range. This is the regime of global navigation satellite systems such as GPS and GLONASS, but also recently of some internet communications systems such as O3B. Satellites in MEO take around 12 hours to orbit the Earth, and each satellite can communicate with a fairly large portion of the Earth's surface. From the perspective of a person on Earth, a spacecraft in MEO will move across the sky much more slowly, and remain in the field of view for an hour.

**Geosynchronous Orbit** (GSO) is at altitude 35,786 km above mean sea level, where the orbital period is matched to Earth's rotation rate about its spin axis. From the perspective of a person on Earth, a spacecraft in GSO will appear as a faint, stationary point source in the sky. GSO satellites can be seen from a large fraction of the Earth's surface. This orbit is traditionally where communications satellites have been placed, including those providing internet or phone services to remote locations. It takes a minimum of 0.24 seconds to send a signal from Earth to a satellite in GSO and back. GSO, in a limited range of orbital inclinations, has long been overcrowded and international regulations restrict its use. With large constellations, we are heading towards similar overcrowding in LEO.

## Anthropogenic space objects

In late 2018, there were around 2,000 active satellites. SpaceX launches have already almost doubled the number of active satellites over the last two years, and all of these in LEO. From published proposals of various companies and states, it seems likely there will be a population of 100,000 or more by the end of the decade [3,4], and a recent filing with the ITU requests 327,000 satellites in a single project. The growth of all tracked anthropogenic space objects (ASOs) is shown in Fig. 1.

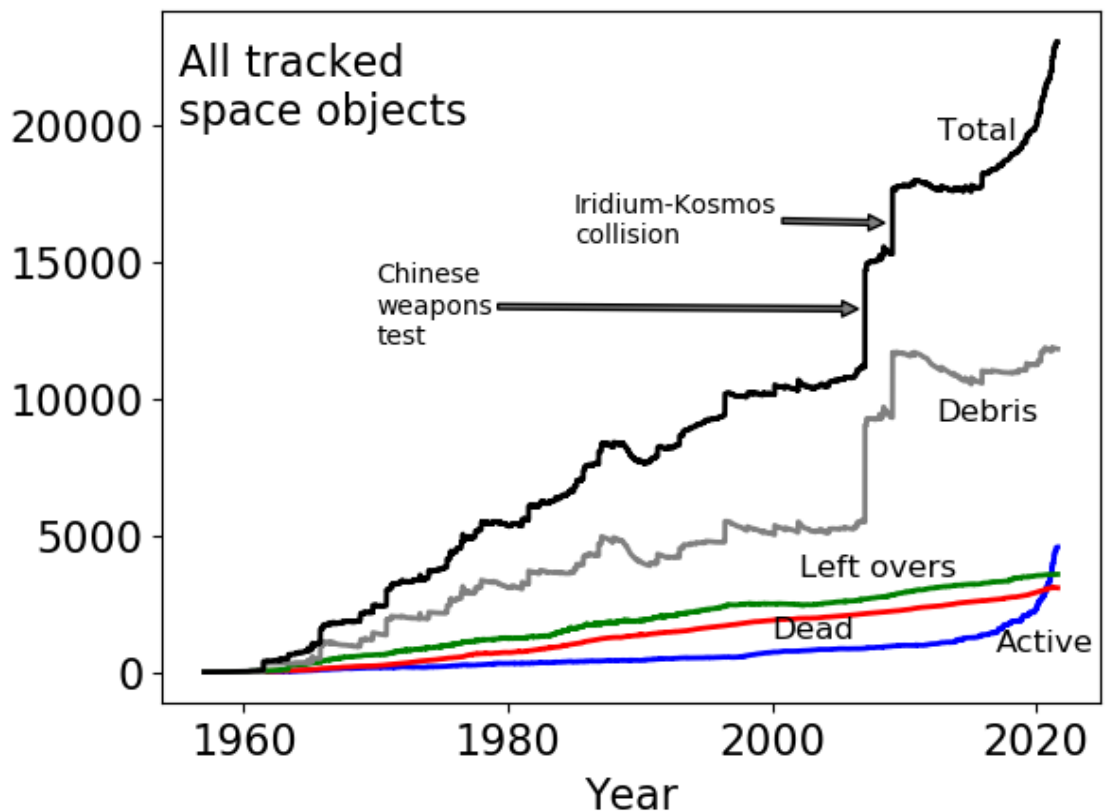

*Fig.1 The growth of all tracked objects in space over time. Updated version of the plot from [3]. Data extracted from the General Catalog of Artificial Space Objects [5].*

Historically, telecommunications satellites were typically placed into GSO. The new satellite constellations, however, are at LEO — partly to minimise the latency (signal delay time), but also to reduce the cost to launch, and ensure rapid decay of failed satellites. Because LEO satellites can access only a small portion of the Earth, *many* more satellites are needed to achieve the equivalent GSO global coverage. The impact of such large constellations has caused considerable disquiet and much work in the astronomical community to mitigate the deleterious effects [3,6,7,8,9,10].

Of the many thousands of satellites that have been launched over the years, most have re-entered, exploded, or continue to orbit the Earth as derelicts, along with other leftover rocket parts. (See Fig. 1.) At LEO a realistic lifespan is about five years, so constellation operators will continuously need to replace satellites. This will require frequent launches and deliberate de-orbiting, leading to a constant turnover within LEO, and the risk of more derelicts from failed satellites. Various processes have led over the years to an ASO population of small pieces of space debris. Down to a size of roughly 10 cm, these can be tracked with telescopes or radar on Earth; there are 22,436 such pieces in The General Catalog of Artificial Space Objects [5].

In Fig. 2 we illustrate the distribution in height and spatial distribution of the entire tracked ASO population, divided into categories. In Fig. 3 we show a distribution of a subset of the tracked ASO population in angular momentum space. These two visualisations make the point that ASOs are not distributed at random, but clustered into orbital shells or highways. Below the size of the tracked objects, the global community hypothesises many untrackable ASOs, possibly as many as 130 million in total. In orbit, typical relative velocities are so high (~10-15 km/s) that even small pieces of debris can cause considerable damage if they collide with something else in orbit (i.e. bullets are small but cause significant damage due to their kinetic energy), and create a growing risk for satellites.

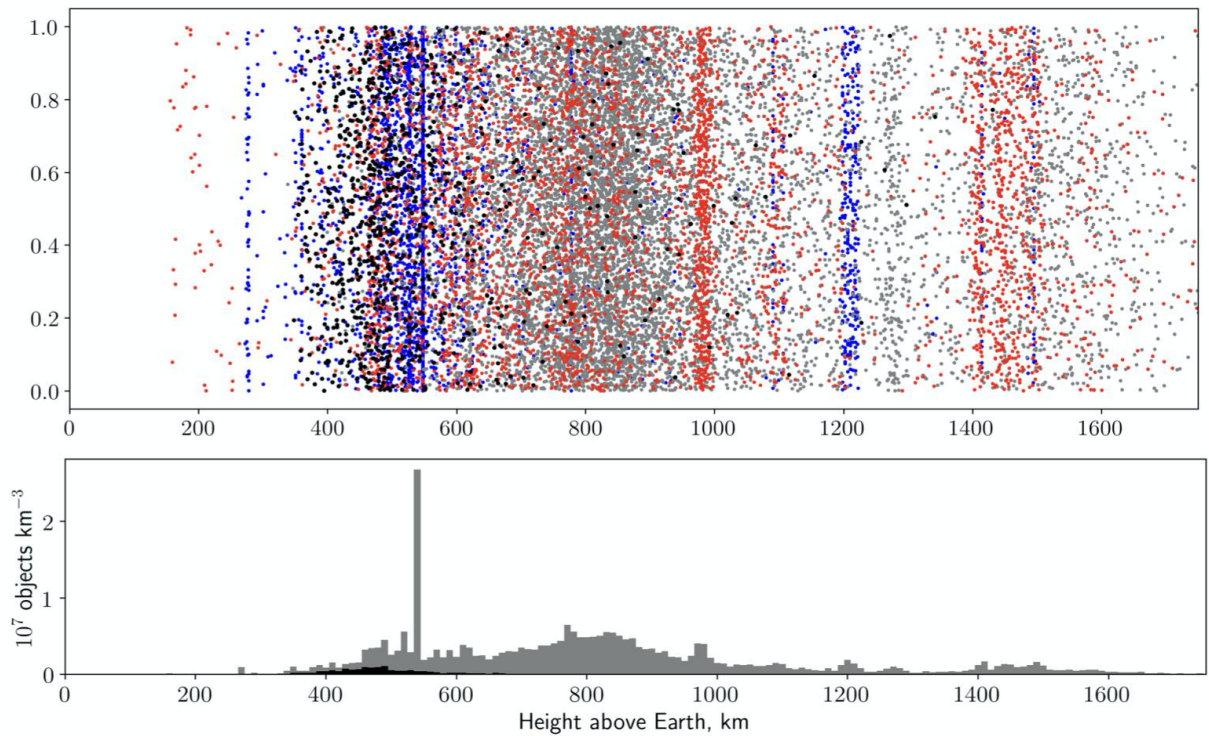

*Fig. 2 Visualisation of the currently tracked objects in Low Earth Orbit. Height is the average of apogee and perigee, relative to a mean Earth radius of 6,378 km. Blue dots are active satellites; red dots are "left overs" such as derelict satellites, rocket bodies and other large parts; grey dots are other debris, down to a scale of approximately 10 cm. Black dots are simulated debris from the recent destruction of Kosmos 1408 by a Russian weapon test, simulated by H.G. Lewis (private communication). The other data points are from the General Catalog of Artificial Space Objects [5]. In the upper plot, the y-axis is simply a random number between 0 and 1, to stretch the points out in two dimensions for clarity. It could be thought of as an artificial azimuth.*

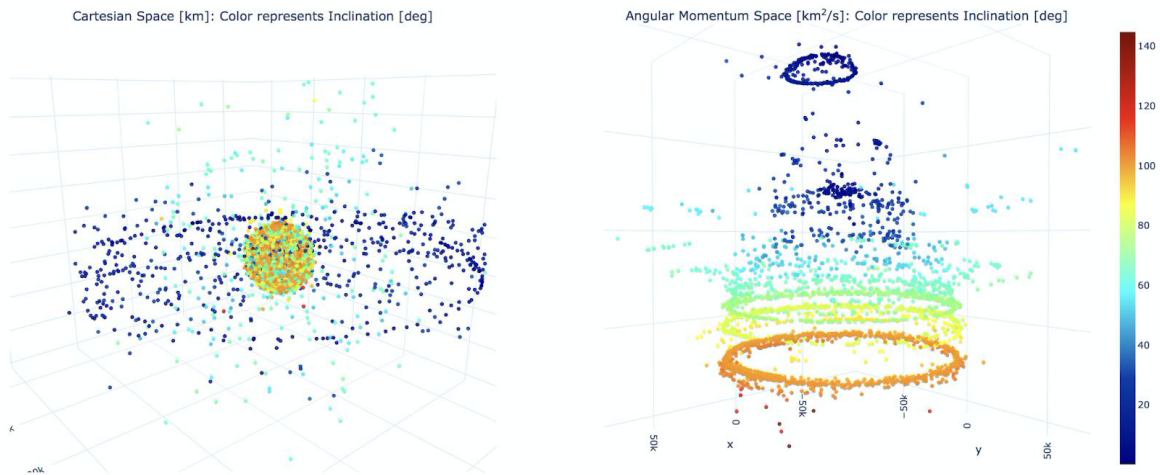

*Fig. 3 Visualisation of 4000 ASOs in various orbital neighbourhoods. The left-hand side plots objects in Cartesian space. The right-hand side plots the same objects in angular momentum space at the same instant. In both cases, the objects are colour coded according to orbit inclination with respect to the Earth's equatorial plane. It can be seen that the ASO population naturally segregates into rings or orbital highways with LEO toward the bottom and GEO as a halo at the top.  Data set from [11].*

Over the years, some non-binding guidelines have emerged to try to minimise proliferation of debris, for example the IADC Space Debris Mitigation Guidelines, and the UN COPUOS LTS guidelines. NASA guidelines [12] state that during active life of satellites, operators should manoeuvre them to avoid collision; and at the end of spacecraft life, it is expected that a spacecraft is either, in the GSO case, moved to a higher storage orbit, or left in a lower orbit where it will decay due to atmospheric drag within 25 years.

# Orbital space and the sky as environments

## Meaning of environment

The Oxford English Dictionary defines environment first as "the surroundings or conditions in which a person, animal, or plant lives or operates". Humans have carried out activities in outer space since 1957, and we have reached a point where these can have deleterious impacts both in space and on Earth's surface. There is therefore a strong case that the concept of environment should extend to orbital space, between 100 km and 36,000 km. The Outer Space Treaty of 1967 [13] and the Liability Convention of 1972 [14] set out general principles quite consistent with this idea, and the LTS guidelines explicitly note that Earth orbit is an environment worth preserving. Definitions of ecosystems could allow space to be included [15].

## The radio interference environment

Satellites communicate with ground stations by radio signals. There is a legal obligation for many bodies such as the US FCC at the national level, and the International

Telecommunications Union (ITU) internationally, to ensure that the activities of operators (including radio astronomers) do not interfere with each other. In this framework, orbital space, and the activities that involve looking through it, are already *implicitly* considered "environment".

## The optical sky as an environment

We can use the same approach/framework for the optical/infrared sky. It is not necessary to be *in* space to be *interacting* with it. The sky constitutes the working environment for astronomy and stargazing, and this inescapably includes orbital space: it can be argued that astronomers carry out space activities in the sense conveyed in Articles IX and XI of the Outer Space Treaty of 1967 [13]. The sky environment has also important cultural significance and has inspired strong traditions around the world since the beginning of human history, such as Maori New Year being associated with the heliacal rising of the Pleiades, or Indigenous Polynesian's star-based navigation [16].

## Orbital space as an environment

Within orbital space itself, the Outer Space Treaty [13] and Liability Convention [14] recognise that the activities of each operator have potential consequences for other operators. However, there is a growing sense worldwide that we should be explicitly considering the *sustainability* of space activities, and considering orbital space as an environment. For example, the recent G7 summit issued a [statement on space sustainability](#) and the World Economic Forum has partnered with the European Space Agency and the University of Texas at Austin, amongst others, to develop a [Space Sustainability Rating](#). Establishing the principle that space activities are subject to environmental laws such as NEPA would be a key step in translating such international good wishes into concrete action.

Many human activities have a locally constrained environmental impact. Most space activities are however inherently global. A satellite may launch from California, but an hour later it is flying over France. Conversely, a Chinese or Russian system will soon appear in the sky over the USA. A coordinated international approach is therefore crucial, but this has to start with each sovereign state recognising its global responsibility. Again, there is a close similarity with other environmental issues such as climate change, or plastic in the sea.

## Cumulative effects and emergent behaviour

The incremental impact of any single proposal for a satellite constellation may be relatively modest, but if all such proposals are allowed because their individual impact is deemed to be modest, the cumulative effect could nevertheless be extremely serious. Furthermore, because of complex interdependencies, the emergent behaviour is not a simple addition and is extremely hard to predict. This is also the case in other environmental issues, such as climate change, and it is widely accepted that environmental assessments need to carefully account for such emergent behaviour. Similar principles should apply to orbital space. Much like other ecosystems, orbital space has a finite "carrying capacity" for traffic. This limit has yet to be globally defined, but it should be evident that if everyone freely populates orbital space without a jointly managed system, this orbital carrying capacity is likely to become saturated, making specific orbital "highways" useless for the safe conduct of space operations and activities. In fact, we should be motivated to define a Space Traffic Footprint, as a Carbon Footprint analogue, that should be loosely interpreted as the burden that any ASO poses on the safety and sustainability of any other ASO and the environment itself.

In order to illustrate potential damage, throughout this article we frequently use a simplified and standardised **potential 2030-era population** of 100,000 objects at an altitude of 600 km. A full environmental assessment would of course use a much more sophisticated approach.

## Impact on astronomy

In considering the impact of satellites on astronomical observations, we have to bear in mind that individual sources of light pollution may be billions or even trillions of times brighter than those that astronomers study, and that many of the most scientifically important observations concern unrepeatable time sensitive or transient events — such as the detection of Near Earth Objects, Supernovae, or Fast Radio Bursts.

### Optical astronomy

ASOs can be seen from Earth because they reflect sunlight. Their brightness depends on numerous factors, such as the size of the satellite, its reflective properties, its height above the Earth and its orientation. As satellites move across the field of view of an astronomical exposure they leave streaks across the image (Fig. 4). For damage already caused by satellites in 2021, see [6,7,8,17] and references therein. To see the likely impact very soon, consider our simplified 2030-era LEO satellite population of 100,000 satellites at a height of 600 km.

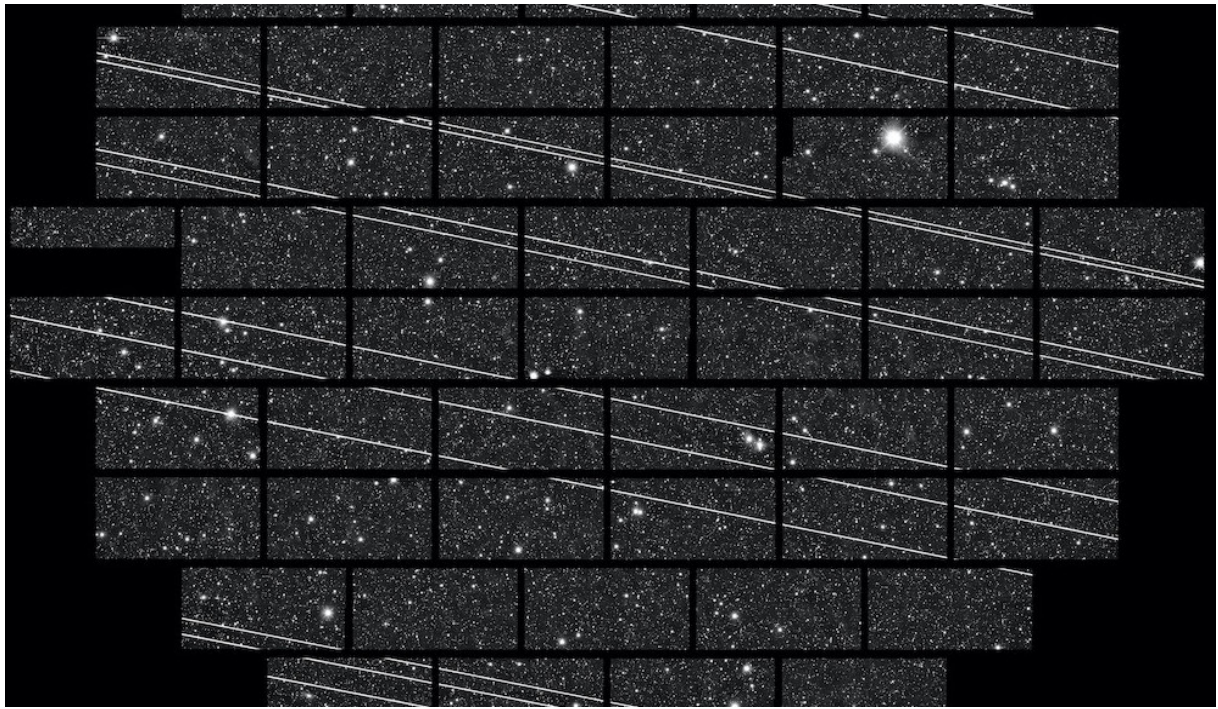

*Fig. 4. An image of the sky taken by the Dark Energy Survey Camera in 2019. Although at that point there were relatively few Starlink satellites, the effect is severe because many*

*Starlink satellites were clumped together during the orbit raising phase shortly afer launch. Credit: CTIO/NOIRLab/NSF/AURA/DECam DELVE Survey.*

Only some satellites are visible above the horizon at a time. For our 2030-era population, roughly 4,300 are above the horizon at any one time, and they cross the sky in about 13 minutes. For a small field of view, there may be only a few percent chance of being affected by a streak, but the observation could be completely wasted and need to be repeated [18]. The more serious impact will be on wide field survey instruments. The Zwicky Transient Facility has already seen an increase in affected images from 0.5% in late 2019 to 18% in August 2021 [17]. The 3.5 degree wide field imager of the Vera C. Rubin Observatory nearing completion in Chile, will contain at least one streak in the majority of exposures [19]. Laboratory experiments using the Rubin Observatory camera detectors show that electronic cross-talk causes streaks to cascade and create additional fainter streaks; this effect can render some scientific analyses impossible because the statistics of the background sky brightness are irrevocably altered. To avoid the crosstalk problem, the satellites would need to be no brighter than 7th magnitude, fainter than the faintest stars visible to the unaided eye at the darkest sites [19].

Furthermore, as an object in space rotates, a brief bright flash or "glint" can occur as a facet or particularly reflective component of the satellite briefly reflects more sunlight to an observer on the ground [20, 21]. For example, Starlink satellites have been seen to change rapidly from fainter than 6th magnitude to almost 3rd magnitude [22]. These extremely bright and short duration (transient) events can mimic some of the most exciting phenomena in modern astronomy. A study in 2020 identified such a flash as the sign of a gamma-ray burst at the edge of the Universe — potentially an extremely exciting discovery. However, a year later it was found that this flash was actually caused by sunlight reflecting off an old Russian Proton rocket part [23]. We do not yet know how frequent this kind of problem will become as the LEO population grows.

When the Earth eclipses a satellite, the satellite is no longer illuminated from the perspective of an observer on Earth. (However, ASOs do emit thermal photons so affect IR sensing even when in eclipse.) As a result, the impact of satellite constellations on astronomical observations is worst near the beginning and end of the night. However, some types of observation simply have to be done at those times; and the fraction of the night affected depends strongly upon the height of the constellation, the geographic latitude of the observatory and the time of year [3,17, 24]. In addition, observations near twilight will see the most streaks, and that is the same time when it is preferable to search for near-Earth objects. As a result, our 2030-era satellite population would yield fewer discoveries of near-Earth asteroids, including ones that may cross Earth's orbit. These are all factors that must be considered carefully in an environmental assessment.

## Radio astronomy

Radio astronomy is affected by satellites using radio signals to relay data back and forth with ground stations and end-user antennas. Detecting faint celestial objects against this anthropogenic background can be potentially very problematic, as the emissions from satellites can be easily a *trillion* times louder than the astronomical targets [3, 4]. In some observations, finely detailed maps are made by combining signals from many interlinked antennas, but the noise problem affects each antenna individually, which physics dictates will always be sensitive to a broad range of directions and frequencies. Unlike optical images, the effect is not a localised streak, but a complex effect across the whole map, which can be hard to recognise and remove — it is like trying to listen to very quiet music in a noisy room. A radio astronomy antenna is sensitive to a range of directions typically a few degrees across — the 'main beam' — but also has reduced sensitivity in very different directions — the "sidelobes". Likewise the satellite antenna emits most of its power in a main beam, but also some in sidelobes. The worst effects, which can potentially damage sensitive electronic receiver systems, are for an alignment between the astronomical and satellite main beams — this rules out radio observations close to GSO targets, and should be avoided even for fast-moving LEO satellites. Sidelobe–sidelobe alignments are much harder to avoid however, as there may be tens or hundreds of LEO satellites in the sky at any one time, and they are all moving quickly across the sky. The net effect is extremely hard to calculate, but a [simulation by the Square Kilometre Array (SKA) project](#) suggests that once the mature Starlink population is in orbit, every observation in the relevant bands will on average take 70% longer.

International regulation of the use of the radio spectrum designates some protected frequency bands for radio astronomy. This approach was originally a great success. However the protected bands were chosen many decades ago when receiver systems were intrinsically narrow-band. Most modern radio astronomy is carried out with state-of-the-art broadband systems, which allow the detection of much weaker natural signals. As a consequence, protection of radio astronomy now relies on geographical radio quiet zones, which some nations provide and some do not. Where available, this zoning can protect against terrestrial interference, but not against satellite interference. While such interference was dominated by a small number of slowly moving GSO satellites, this was acceptable, but the new LEO constellations could lead to very serious issues. The new systems inevitably overlap with satellite communication bands. Furthermore, volume manufacture and deployment of large numbers of relatively low-cost satellites is likely to increase the chance of sideband leakage into protected bands.

Like the spatial interference issue, the assignment of protected bands sets a precedent, as frequency interference is implicitly recognised as an environmental effect. Recognising that the issues should be subject to environmental laws such as NEPA is the logical next step as the problems get much worse.

## Space astronomy

Some spacecraft used for astronomy are placed at very large distances from the Earth, and are not affected by LEO satellites. Many however, like the Hubble Space Telescope (HST), are in LEO, and can certainly suffer from streaking. Occasionally a satellite may pass relatively close by (< 100 km) in which case the streak caused is an extremely bright out of focus stripe, obliterating a significant fraction of the image. An example is shown in Fig. 5. A recent study [25] showed that, depending on the instrument and observational parameters being used, between 2% and 8% of HST images were affected by satellite streaks, but also that the frequency was changing with time, reflecting the growth of the LEO satellite population. Our 2030-era population indicates that by the end of the decade a third of HST images will be affected, and likewise future LEO-based science missions, such as the Xuntian wide field observatory being built for the Chinese Space Station.

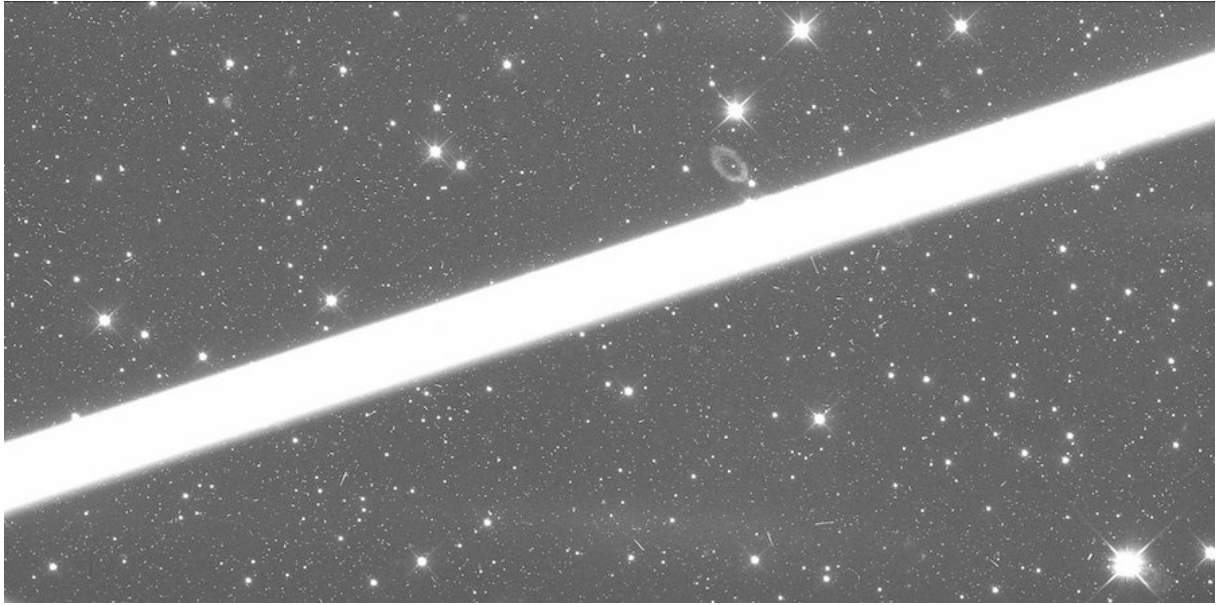

*Fig.5. An observation made using the Hubble Space Telescope in November 2020. The streak seems likely to have been made by Starlink 1619, only a few km above Hubble, thus creating a wide out of focus trail. (Image credit: Mikulski Archive for Space Telescopes (MAST). Science PI: Simon Porter.)*

## Mitigations, damage and their cost

The international astronomical community has had multiple meetings to discuss how to address the new landscape of increasing numbers of bright LEO satellites, leading to key reports [6–10]. A report by the US Government independent advisory body JASON was also commissioned by the National Science Foundation [4]. Astronomers have engaged with satellite companies to discuss ways to mitigate the problems. For optical astronomy, this has included ideas such as painting satellites black, changing their orbits and orientations, adding sun visors, and providing detailed positions and trajectories so that observatories can avoid pointing at them. For radio astronomy, key mitigations further include redirecting beams away from major observatory facilities and employing sophisticated signal filtering. None of these mitigations can fully avoid LEO satellite constellations harming astronomical science however [7, 8, 10]. Launching significantly fewer satellites is the only mitigation that could do this.

The consequences of the current and proposed growth of satellite constellations have a direct cost from repeating or extending observations, wasting scientist time, and even negatively affecting their careers. Implementing mitigations will also impose significant costs, either on the astronomical community (and so the taxpayer), or on the satellite operator companies, or on both. We do not attempt to assess those costs here. Rather, we point out that this is a classic example of environmental damage, externalising true costs. To give one example, one significant conclusion from the SATCON2 Observations Working Group [8] is to establish a coordinated satellite observation hub under the umbrella of a larger International Astronomical Union (IAU) [Centre for the Protection of the Dark and Quiet Sky from Satellite Constellation Interference](). Such a long-term mitigation activity will require significant sustained resources.

We note that the US FCC order under current legal discussion has, quite correctly, encouraged SpaceX to continue engagement with the astronomical community. However, these productive collaborations ought to proceed within the context and guidance of Environmental Assessment.

## Impact on public access to sky

A more complete discussion of this topic can be found in the Community Engagement Working Group report from SATCON2 [9], but it is worth restating the main points here.

### Public access to the stars

IAU's Resolution B5, "[In Defence of the Night Sky and the Right to Starlight]()" (2009), asserts that "'[a]n unpolluted night sky that allows the enjoyment and contemplation of the firmament should be considered a fundamental socio-cultural and environmental right."

A greatly increased number of satellites can significantly alter our whole perception of the night sky in the long-term, appearing as "fake stars"; according to our model 2030-era population of 100,000 bright satellites at 600 km, the number of visible fake stars could well rival *the number of visible real stars* [3,24]. They will be towards the fainter end of what one can see with the unaided eye, affecting especially the remaining uncontaminated places to observe the sky, even for the whole night (depending on seasons and latitudes). Even so, a significant number of satellites at the margins of visibility may create an unsettling effect of constant wriggling and squirming.

In addition, for many Indigenous people, the night sky is an active and vital part of culture, storytelling and inheritance from one generation to another. It is reasonable to claim that access to the night sky environment, including unobstructed views of the stars, can be considered a basic human right for all people. Satellites will also significantly affect amateur astronomy and citizen science, which have become relevant particularly in recent years as an integral part of scientific exploration. For a typical 7 degree binocular field of view, taking our model 2030-era population, around eight satellites will be visible everywhere you look and they will typically be the brightest objects in the field of view. They will move across the field of view in about ten seconds, continuously being replaced by new ones. Meanwhile, many amateur astrophotographers will suffer the same problem as professional astronomers — streaks in most of their images.

## Collision impacts on space operations

### Growth of space debris

The space community loosely divides objects in space into 'active satellites', 'dead and leftovers' such as derelict satellites and rocket stages, and 'debris', resulting from fragmentations, explosions and collisions. As described above, debris can in turn be divided into 'tracked debris', down to 10 cm size, and smaller 'untracked' debris, which can only be estimated. The number of debris objects grows faster with time than the leftover population, and specific events like the Iridium–Kosmos collision in 2009, and anti-satellite weapons tests like those conducted by China, the USA, India and Russia in 2007, 2008, 2009, 2019, and 2021 respectively, can cause large leaps (Fig.1).

A concern arising from these trends is that certain orbital highways will exceed their carrying capacity, rendering them unusable. This saturation would become manifest when our decisions and actions can no longer prevent the loss, disruption, or degradation of space operations, services and activities. When we launch dozens of satellites every few weeks, we remove the environment's ability to inform us of the unintended consequences of our actions and we cannot predict what the dynamic equilibrium state actually is. To wit, it clouds our decision intelligence.

## Classifying collisions

We can roughly classify collisions into *minor*, *disabling* and *disrupting or lethal*. Anything 1 mm in size or larger can cause minor damage, such as perforating a solar array. This can include natural micro-meteoroids as well as satellite debris. A piece of debris 10 cm in size will have a mass of about 1 kg, and if it is moving at 10 km/s (typical for relative velocities in LEO), it can *completely destroy* an active satellite [4]. Between these extremes, a 1 cm piece of debris is capable of disabling an active satellite [4]. Note that pieces as small as 1cm are not currently tracked, and even very small pieces of debris or micrometeoroids can cause damage, as seen in the recent case of an impact on the Canadian robotic arm of the International Space Station.

## Risk of disruptive (lethal) collisions

Without avoidance methods, the current debris density means there will on average be one collision per satellite every 50 years in LEO, with a piece of debris that is 10 cm or larger [4]. However, large objects are tracked and orbital elements made publicly available, so that potential collisions can be predicted and actively avoided. 'Conjunctions', where one satellite passes within a few km of another, happen many times every year, but so far only one major accidental collision has taken place. The presence of large constellations will increasingly put any avoidance manoeuvring system under severe stress, with some close calls summarised and analysed in [3, 4].

## Risk of disabling collisions

Calculating the likelihood of *disabling* damage by debris with size > 1 cm to active satellites is a complex problem requiring many physical variables. Note that disabling a satellite leaves an uncontrolled derelict that may then be a danger to other spacecraft. Simplified modelling of the possible future is given in the recent comprehensive JASON report [4]. This includes allowance for continually de-orbiting satellites at the end of an assumed 5 year lifetime. For a target population of 10,000 active satellites, debris grows only slowly but we can expect about 300 disabling collisions within the next 30 years (Figure 6a). For a target population of

40,000 satellites, debris growth is dramatic and there will be hundreds of disabling collisions within a few years (Figure 6b). After a few decades, it is likely that satellites will be disabled faster than they are launched. These calculations were performed for a Starlink population, and a similar calculation was carried out for the OneWeb constellation at 1,200 km. The results were subtly different but equally disturbing.

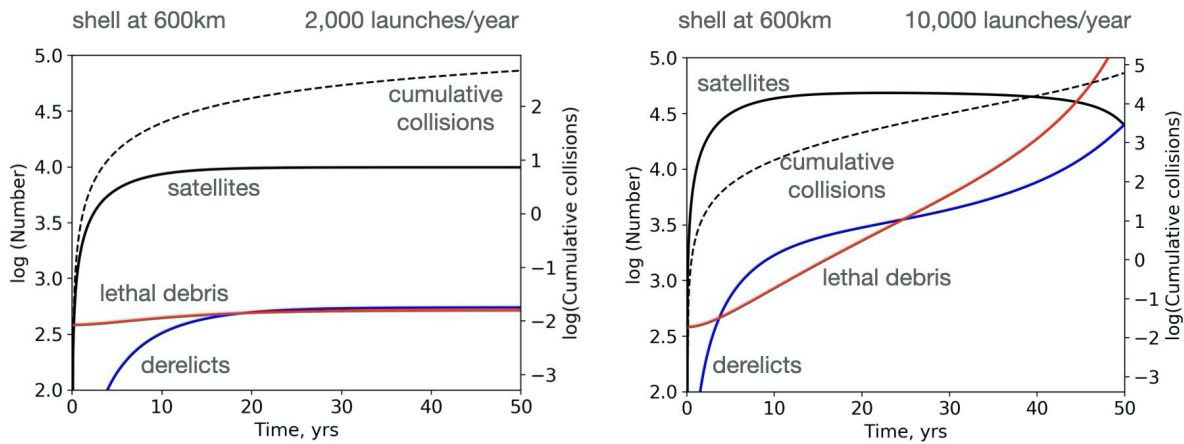

*Fig. 6. The evolution of the satellite population, debris population and cumulative collisions for two possible satellite constellation scenarios at a height of 600 km with frequent de-orbiting. a. 2,000 launches per year aimed at a stable population of 10,000 satellites. b. 10,000 launches per year aimed at a population of 40,000 satellites. Calculations made by authors using the JASON model [4], and using the same parameters as those in Fig. 20 of that report.*

Further modelling [26] looks more specifically at the collision rate likely in the de-orbiting zone, and finds that even with current debris density, each Starlink satellite has a roughly 50% chance of a collision each year from untracked debris. This collision probability would rise dramatically with any increase in debris.

## Brief statements on other potential impacts

### Atmospheric pollution

Atmospheric effects are discussed by Boley & Byers [26]. All rocket launches result in emissions with negative impacts on the atmosphere, including $CO_2$, $NO_x$, soot and $H_2O$ in the mesosphere. So far these are minor contributors to the global budget, but the huge number of launches required to build and maintain constellations of thousands of satellites will increase pollution by a large factor. Future rocket types may also deposit other materials that could increase global warming directly in the stratosphere. Re-entering satellites and debris also deposit fine particulates during their burn-up. In particular, aluminium will be deposited at a rate that exceeds that from naturally entering micro-meteoroids, and may have an effect on the Earth's albedo. Ongoing climate change may also alter thermospheric density enough to significantly increase orbital decay lifetimes in LEO [27].

### Ground and airspace collision

It is unlikely that all de-orbited satellites will burn up completely, or that all surviving rocket parts, including unspent fuel, will be successfully dumped in the sea, so damage to property and even life will be an increasing risk. Disposing of satellite remnants in a marine environment has environmental risk, which has been successfully challenged in the past [28,29]. The risk to life and impact on the environment is non-trivial. Based on a population of 16,000 satellites in LEO, it has been estimated that by 2030 the probability of casualties on the ground will rise to 0.1/year (presentation to UN COPUOS committee, quoted in [30]). Descending debris also poses a risk to aircraft. From the same UN presentation, predictions suggest a 1 in a 1000 chance of an aircraft being struck each year, but with some 300 passengers per aircraft, that means 0.3 casualties per year. A possible population of 100,000 satellites increases the casualty rate by many times. The first aircraft strike or ground casualty is only a matter of time.

### Animal and plant ecosystems

Numerous animal species ranging from insects to mammals to birds are known to orient themselves during migration and foraging activities using the stars and the Milky Way [31–34]. Roughly 40% of bird species migrate, and roughly 80% of those migrating species migrate at night, many of them using the stars to navigate [35,36]. While we cannot yet know whether those species will be sensitive to many additional "stars" appearing to move rapidly across the sky, reasonable predictions of potentially significant harm are already appearing in the scientific literature [37]. It is also possible that integrated sky brightness may increase significantly, with further disruption to some species and ecosystems. [38, 7, Bio-Environment Report].

### Space Weather issues

Activity from the Sun, called space weather, has dramatically affected satellites in the past. Charged particles are ejected from the Sun at high speeds during solar storms, and these charged particles can have negative effects on the on-board electronics in satellites, causing them to temporarily shut down in a "safe mode" until a reset command can be issued from the ground. Satellites can even have their electronics overloaded and be permanently disabled. With the huge increase in the number of satellites and the increased collision risk, active collision avoidance by many satellites will be frequent. If satellites are disabled, even temporarily, they will lose the ability to manoeuvre around hazards and the collision risk will increase dramatically every time a satellite enters "safe mode" or is disabled. The frequency and intensity of solar storms varies in an eleven-year cycle and the next Solar Maximum, when solar activity will be at its peak, is predicted to be in 2024-2025. The population of satellites by then is expected to be several times higher than it is today and it is worth noting that a relatively minor geomagnetic storm resulted in [an unexpected descent and burnup](#) of 40 Starlink satellites in February 2022.

## Conclusion

We have laid out the argument that there is an urgent need for orbital space to be considered part of the human environment. Adequately addressing the problems detailed above will require a holistic approach that treats orbital space as part of the environment, and worthy of environmental protection through existing and new policies, rules and regulations at national and international levels. We urge decision- and policy-makers to consider the environmental impacts of all aspects of satellite constellations, including launch, operation and de-orbit, and to work with all stakeholders to co-create a shared, ethical, sustainable approach to space.

**Acknowledgements**

We are grateful to H.G.Lewis at the University of Southampton for providing a dataset of orbital elements of simulated debris from the recent C1408 ASAT event, as well as for general discussions on the topic of space environmentalism. Many further colleagues have contributed indirectly to this article - through comments on the open document during August 2021 in preparation for the Amicus Brief, and in general discussions at astronomical meetings during 2020 and 2021 at which the issue of the impact of constellations was discussed.

**Author Information**

**Affiliations**

Andy Lawrence
**Institute for Astronomy, School of Physics and Astronomy, University of Edinburgh, UK. Scottish Universities Physics Alliance (SUPA)**

Meredith L. Rawls
**Department of Astronomy and DiRAC, University of Washington/Vera C. Rubin Observatory, Seattle, USA**

Moriba Jah
**Associate Professor Aerospace Engineering & Engineering Mechanics**
**The University of Texas at Austin; and Co-Founder and Chief Scientist, Privateer Space Inc.**

Aaron Boley
**Department of Physics and Astronomy, University of British Columbia, Vancouver, Canada**

Federico Di Vruno
**SKA Observatory, Jodrell Bank, Manchester, UK.**

Simon Garrington
**Jodrell Bank Observatory, University of Manchester, UK**

Michael Kramer

**Max-Planck-Institut fuer Radioastronomie, Bonn, Germany; and Jodrell Bank Centre for Astrophysics, University of Manchester, Manchester, UK**

Samantha Lawler
**Campion College and the Department of Physics, University of Regina, Canada**

James Lowenthal
**Smith College, Northampton, Massachusetts, USA**

Jonathan McDowell
**Center for Astrophysics, Harvard and Smithsonian, USA.**

Mark McCaughrean
**European Space Agency, Noordwijk, The Netherlands.**

**Corresponding Author**

Correspondence to [Andy Lawrence](Andy Lawrence)

**Contributions**

The article was conceived, initiated and led by AL, who wrote the initial draft, in collaboration with MR and MJ, who also wrote major parts of the text. All other authors contributed significant text or key technical or scientific points.

**Competing interests**

The authors declare no competing interests.

**Supplementary Information**

The datasets and Jupyter notebooks used in the construction of Figures 1, 5 and 6 are available at https://github.com/andyxerxes/Space-environment-paper. The datasets used for Fig. 2 are available at https://doi.org/10.18738/T8/LHX5KM